# Molecular center photoabsorption in solid Ar

A.N. Ogurtsov,[1,2] E.V. Savchenko,[1] S. Vielhauer[3] and G. Zimmerer[3]

[1]Institute for Low Temperature Physics and Engineering of NASU, Lenin Avenue 47, 61103 Kharkov, Ukraine
[2]National Technical University "KhPI", Frunse Street 21, 61002 Kharkov, Ukraine
[3]Institut für Experimentalphysik der Universität Hamburg, Luruper Chausee 149, 22761 Hamburg, Germany

The relaxation of electronic excitations in an easily deformable lattice of rare gas solids is accompanied by pronounced trapping effects. A remarkable feature of the trapping of excitons and holes in *atomic* cryocrystals is a creation of *molecular*-like neutral $R_2^*$ and ionic $R_2^+$ centers in the course of energy relaxation. The radiative decay of self-trapped excitons (STE) and holes (STH) in solid Ar, Kr, Xe forms various molecular luminescence bands, which are mainly in a vacuum ultraviolet region [1] and commonly used to investigate the dynamics of electronic and elementary lattice processes. The most prominent feature in the low temperature luminescence from atomic cryocrystals is the so-called *M*-band (molecular luminescence band) which is formed by transitions in molecular STE $R_2^*$. The radiative transitions in self-trapped holes $R_2^+$ produce near-ultraviolet *H*-bands. After recombination of $R_2^+$ with an electron it also turns into the neutral molecule $R_2^*$ Irrespective of the manner of their formation, the decay of excimer molecules is the final stage of energy relaxation and it provides important information about energy pathways for relaxation of electronically deposited energy. In the context of relaxation processes in atomic cryocrystals two questions are particularly interesting: (i) Whether molecular centers are formed only as a result of trapping of electronic excitations, or they exist in the lattice prior to excitation of the samples? (ii) What is the spatial distribution of molecular trapping centers in the volume of the crystal? The recent analysis of the molecular spectra under different excitation conditions, excitation energies and crystal growth conditions made it possible to elucidate the internal structure of the *M*-bands. Each of the *M*-bands can be well approximated by two Gaussians: low energy subband $M_1$ and high energy one $M_2$. The subband $M_2$ is dominant in the luminescence of more perfect samples. The spectra of samples with a great number of initial defects are determined mainly by the "defect'" component $M_1$. This suggests that the subband $M_2$ is emitted by the excitons which are self-trapped in the regular lattice while the component $M_1$ is during trapping that occurs with the lattice imperfections involved.

In solid Xe and Kr, using time-resolved fluorescence spectroscopy under selective photoexcitation by synchrotron radiation, the threshold energy $E_{thr}$ of photoabsorption by molecular trapped stats was detected in excitation spectra of free-exciton luminescence [2]. This energy separates the range of photoexcitation where free and self trapped excitons coexist from the range where photon absorption creates only free excitons. In solid Ar the free exciton photoluminescence has not been detected till now. Nevertheless it is possible to control the creation of primary and secondary free-excitons by luminescence of molecular self-trapped excitons [3]. In the present work we search for $E_{thr}$ in solid Ar using molecular luminescence.

The experiments were performed at the SUPERLUMI experimental station at HASYLAB, DESY, Hamburg. Selective photon excitation was performed with $\Delta\lambda$=0.2 nm. The luminescence was spectrally dispersed by 0.5 m Pouey monochromator with $\Delta\lambda$=2 nm equipped with multisphere plate detector. The time-window technique was used and signal was measured within a time window (length $\Delta t$) correlated with the excitation pulse (delayed by $\delta t$).

Figure 1 shows the reflection (a) of solid Ar in the photon range 11.2–12.7 eV; the excitation spectrum of *W*-band (b) measured at $h\nu$=11.27 eV; photon yield (c) from the sample measured at $h\nu$=11.47 eV in the "long" time window ($\delta t$=10 ns, $\Delta t$=35 ns); and the excitation spectrum of *M*-band measured at $h\nu$=9.38 eV (predominant emission of "defect" $M_1$-subband). All spectra show the threshold energy $E_{thr}$=11.63 eV. Starting from this energy the photoabsorption by molecular centers begins well below the bottom of the lowest $\Gamma(3/2)$ n=1 excitonic band ($E_{FE}$=12.06 eV). Thus, in solid Ar the direct population of molecular emitting centers occurs like in the case of solid Xe and Kr [2]. The low-energy extension of the photoabsorption by molecular centers below the bottom of the $\Gamma(3/2)$ n=1 excitonic band growth with decreasing the atomic mass and amounts to 0.18 eV for Xe, 0.3 eV for Kr and 0.43 eV for Ar.

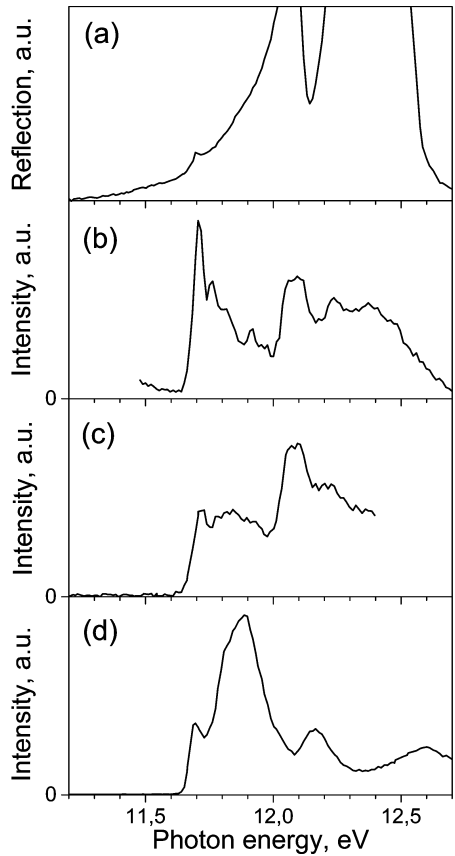

Figure 1: (a) – Reflection of solid Ar. (b) – Excitation spectrum of *W*-band. (c) – Spectrum of photon yield measured at $h\nu$=11.47 eV. (d) – Excitation spectrum of *M*-band. All data were measured at *T*=10 K.

*Acknowledgements*. The support of the DFG grant 436 UKR 113/55/0 is gratefully acknowledged.